\documentclass[a4paper,twoside]{article}
\newcommand{\affil}[1]{$^{\rm #1}$}
\newcommand\ion[2]{#1$\;${\scshape{#2}}}

\usepackage[authoryear]{natbib}
\usepackage{amssymb}
\usepackage{cancel}
\bibpunct{(}{)}{;}{a}{}{,}
\usepackage{graphicx}
\date{} 
\title{\large\bf\flushleft SN\,2011fe: A Laboratory for Testing Models of Type Ia Supernovae}
\author{\parbox{\textwidth}{\flushleft
\vspace{-0.5cm}
{\it Laura Chomiuk\affil{A,B,C}}\\
\vspace{0.4cm}
{\small \affil{A}\,Department of Physics and Astronomy, Michigan State University, East Lansing, MI 48824}\\
{\small \affil{B}\,National Radio Astronomy Observatory, P.O. Box O, Socorro, NM 87801}\\
{\small \affil{C}\,Email: chomiuk@pa.msu.edu}}}
\begin{document}
\twocolumn[
\begin{changemargin}{.8cm}{.5cm}
\begin{minipage}{.9\textwidth}
\vspace{-1cm}
\maketitle
%
%
\small{\bf Abstract:}
SN\,2011fe is the nearest supernova of Type Ia (SN~Ia) discovered in the modern multi-wavelength telescope era, and it also represents the earliest discovery of a SN~Ia to date. As a normal SN~Ia, SN\,2011fe provides an excellent opportunity to decipher long-standing puzzles about the nature of SNe~Ia. In this review, we summarize the extensive suite of panchromatic data on SN\,2011fe, and gather interpretations of these data to answer four key questions: 1) What explodes in a SN~Ia? 2) How does it explode? 3) What is the progenitor of SN\,2011fe? and 4) How accurate are SNe~Ia as standardizeable candles? Most aspects of SN\,2011fe are consistent with the canonical picture of a massive CO white dwarf undergoing a deflagration-to-detonation transition. However, there is minimal evidence for a non-degenerate companion star, so SN\,2011fe may have marked the merger of two white dwarfs. 


\medskip
\medskip
\end{minipage}
\end{changemargin}
]
\small

\section{Introduction}

\begin{figure*}[]
\begin{center}
\includegraphics[scale=0.8, angle=0]{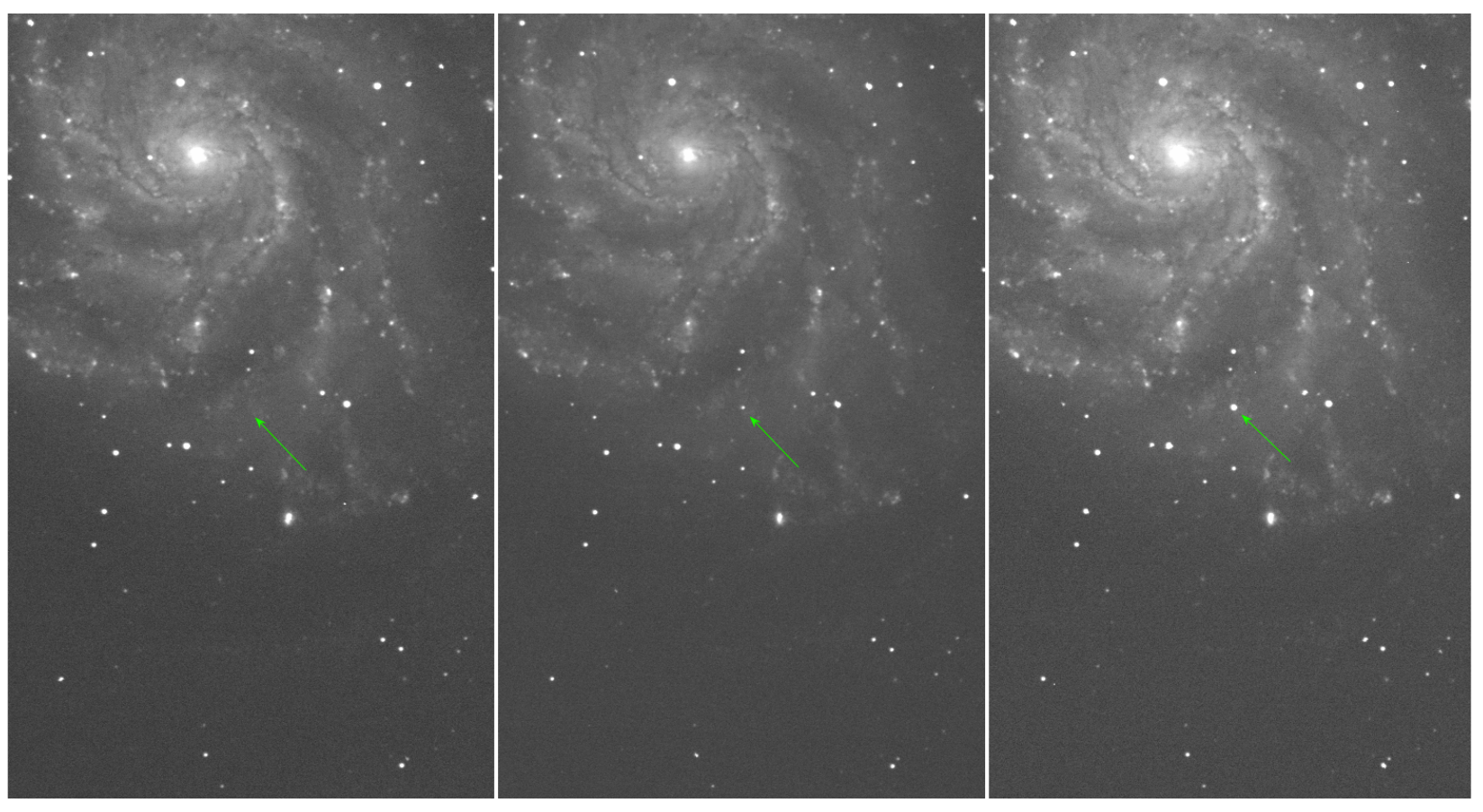}
\vspace{-0.4cm}
\caption{Images of M101 obtained on three successive nights, from left to right: 2011 Aug 23.2, Aug 24.2, and Aug 25.2 UT. The green arrow points to SN\,2011fe, which was not detected in the first image but subsequently brightened dramatically. Figure from \citet{Nugent_etal11b}, reprinted by permission from Macmillan Publishers Ltd: Nature, copyright 2011.}\label{nugent11fig}
\end{center}
\end{figure*}

Discovered on 2011 August 24 by the Palomar Transient Factory, SN\,2011fe\footnote{The source was originally dubbed PTF\,11kly.} was announced as a Type Ia supernova (SN~Ia) remarkably soon after explosion (just 31 hours; \citealt{Nugent_etal11a, Nugent_etal11b}). SN\,2011fe is nearby, located in the well-studied galaxy M101 at a distance of $\sim$7 Mpc (Figure \ref{nugent11fig}; \citealt{Shappee_Stanek11, Lee_Jang12}). As the earliest and nearest SN~Ia discovered in the modern multi-wavelength telescope era, SN\,2011fe presents a unique opportunity to test models and seek answers to long-standing questions about SNe~Ia.

Such a testbed has been sorely needed. SNe~Ia are widely used by cosmologists to measure the expansion parameters of the Universe, and led to the Nobel Prize-winning discovery of dark energy \citep{Riess_etal98, Perlmutter_etal99}. However, important unknowns remain that stymie the use of SNe~Ia as precise cosmological tools. The progenitor systems of SNe~Ia are poorly understood, and the explosion mechanism itself is elusive (reviews by \citealt{Hillebrandt_Niemeyer00, Livio01, Howell11, Wang_Han12} contain more details). 


This paper reviews the significant body of work already published on SN\,2011fe as of mid-2013. I focus on five questions: (i) Is SN\,2011fe a normal SN~Ia?  (ii) What exploded in SN\,2011fe? (iii) How did it explode? (iv) What is the progenitor of SN\,2011fe? and (v) How accurately can we use SNe~Ia as standard candles? A concise summary follows in Section 6.

\section{SN\,2011fe: A normal SN~Ia}

\begin{figure}[h]
\begin{center}
\includegraphics[scale=0.8, angle=0]{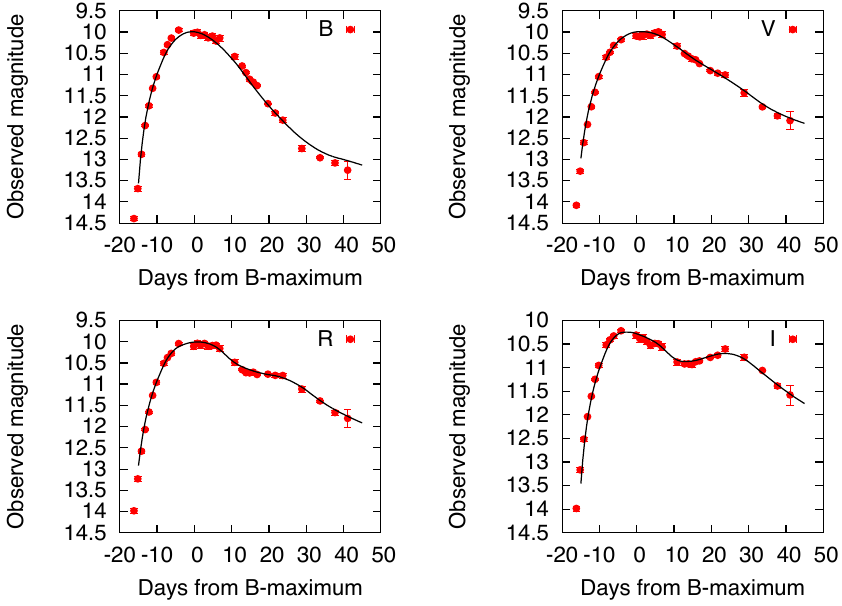}
\vspace{-0.4cm}
\caption{$BVRI$ Light curves for SN\,2011fe measured in the Johnson-Cousins system (Vega magnitudes). Best-fit SN~Ia templates from SALT2 are plotted as black lines. Figure from \citet{Vinko_etal12}, reproduced with permission \copyright ESO.}\label{vinko12fig}
\end{center}
\end{figure}

The multi-band light curve of SN\,2011fe, measured in exquisite detail at UV through IR wavelengths, is typical of SNe~Ia (Figure \ref{vinko12fig}; \citealt{Vinko_etal12, Brown_etal12, Richmond_Smith12, Munari_etal13, Pereira_etal13}). In the $B$-band, the light curve declines by $\Delta m_{15} = 1.1$ mag in 15 days (\citealt{Richmond_Smith12, Pereira_etal13}). To power the light curve of SN\,2011fe, $\sim$0.5 M$_{\odot}$ of $^{56}$Ni is required (\citealt{Nugent_etal11b, Bloom_etal12, Pereira_etal13}). This $^{56}$Ni mass is quite typical for SNe~Ia \citep{Howell_etal09}.

In addition, time-resolved optical spectroscopy shows SN\,2011fe to be a spectroscopically-normal SN~Ia \citep{Parrent_etal12, Pereira_etal13, Mazzali_etal13}.  SN\,2011fe can be classified as ``core normal" in the schemes of \citet{Benetti_etal05} and \citet{Branch_etal06}.

In all relevant details, SN\,2011fe appears to be a normal SN~Ia. It has, therefore, been taken to be representative of its class. Conclusions reached for SN\,2011fe may be extrapolated to SNe~Ia generally, but we must be careful in this extrapolation, remembering that SN\,2011fe is only one object. A number of recent studies have shown that SNe~Ia are diverse, and that differences in their observational properties may imply real variety in progenitor systems and explosion mechanisms \citep[e.g.,][]{Foley_etal12, Wang_etal13}
.

\section{What exploded in SN\,2011fe?}\label{what}

While it has long been thought that a SN~Ia marks the complete disruption of a white dwarf \citep[e.g.,][]{Hoyle_Fowler60}, little direct proof existed until SN\,2011fe.


\subsection{Radius of the exploded star} \label{radius}
The radius of an exploding star (R$_{\star}$) can be estimated through the phenomenon of shock breakout: when the SN shock emerges from the surface of the star, it produces a distinctive photometric signature. Shock breakout is expected to appear as an early-time excess in the light curve, with the luminosity and duration of the excess scaling with the radius of the exploding star \citep{Rabinak_Waxman11, Kasen10, Piro_etal10}.

\begin{figure}[t]
\begin{center}
\includegraphics[scale=0.9, angle=0]{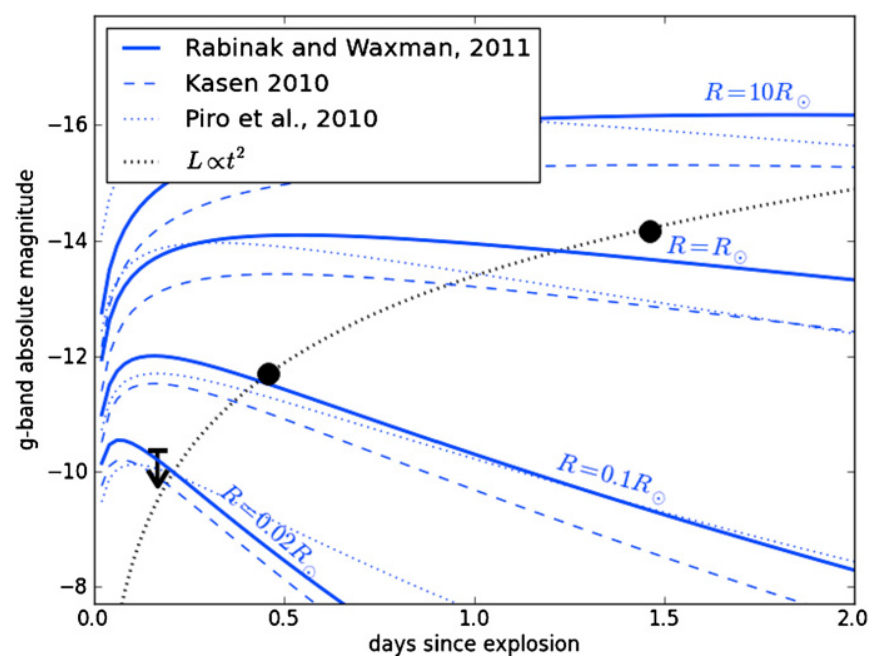}
\vspace{-0.5cm}
\caption{Early-time optical light curve for SN\,2011fe plotted as black dots. The light curve is well fit with a simple $L \propto t^2$ power law (dotted black line). Three variations of shock breakout models are plotted as blue lines \citep{Rabinak_Waxman11, Kasen10, Piro_etal10}, and are shown for different radii of the exploding star. Assuming that the explosion time can be determined from the simple power-law fit \citep{Nugent_etal11b}, the non-detection was obtained four hours after explosion and  implies a size R$_{\star} \lesssim 0.02$ R$_{\odot}$ for the exploding star. Figure from \citet{Bloom_etal12}, reproduced by permission of the AAS.}\label{bloom12fig}
\end{center}
\end{figure}
\begin{figure}[t]
\begin{center}
\includegraphics[scale=0.85, angle=0]{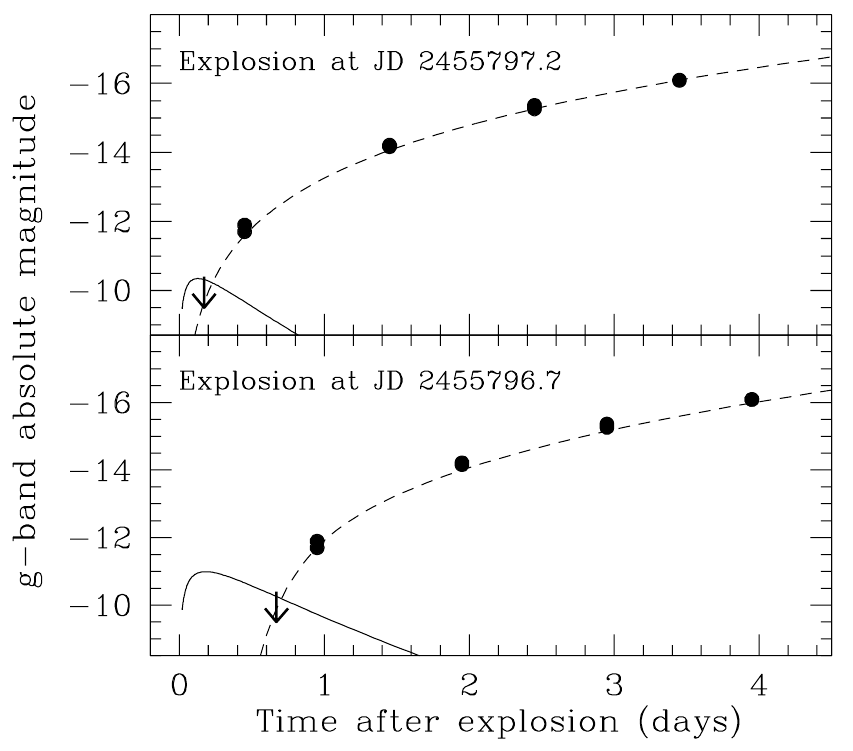}
\vspace{-0.4cm}
\caption{The limits on a shock breakout signature and R$_{\star}$ depend on the estimated explosion date. In both panels, the measured early-time light curve of SN\,2011fe is shown as filled circles and an upper-limit arrow. The shock breakout signature is a sold black line, and the SN light curve, powered by $^{56}$Ni, is a dashed line. The top panel shows the model of \citet[][as in Figure~3]{Bloom_etal12}, assuming an explosion date of UT 2011 August 23 16:30 and constraining R$_{\star} \lesssim 0.02$ R$_{\odot}$. The bottom panel is for an explosion date of UT 2011 August 23 4:30, yielding a 12-hour long dark phase and a larger radius constraint R$_{\star} \lesssim 0.04$ R$_{\odot}$. Figure from \citet{Piro_Nakar12}, reproduced by permission of the authors.}\label{piro_nakar12fig}
\end{center}
\end{figure}

The shock breakout constraint depends crucially on knowing the precise time of the explosion. This is estimated to be UT 2011 August 23 16:$29\pm20$ minutes by \citet{Nugent_etal11b}. They derive this time by fitting a power law to the early optical light curve, following the expectation that the SN luminosity $L$ will increase as the area of the optically-thick photosphere, producing the relation $L \propto t^2$. This simple model fits the photometry very well over the first four days (see Figures \ref{bloom12fig} and \ref{piro_nakar12fig}).

The modeling of the shock breakout is aided by the serendipitous availability of optical imaging of M101 that had been obtained 
a mere four hours after Nugent et al.'s estimated time of explosion, but before the SN was actually discovered \citep{Bloom_etal12}. These data, obtained with The Open University's 0.4-m telescope, yield a robust non-detection at this epoch. The first detection of SN\,2011fe was made 11 hours after Nugent~et~al.'s estimated explosion time. 

The faint optical flux at very early times places strong constraints on the shock breakout signal, implying that the exploding star was compact, with a radius R$_{\star} \lesssim 0.02$ R$_{\odot}$ (Figure \ref{bloom12fig}). From the measured $^{56}$Ni mass, we know the star was $\gtrsim$0.5 M$_{\odot}$. Only degenerate stars satisfy these mass and radius constraints. Thus the exploded body in SN\,2011fe must have been a neutron star or white dwarf. There are no plausible mechanisms for producing SN~Ia-like thermonuclear yields from a neutron star \citep[e.g.,][]{Jaikumar_etal07}. Radius constraints therefore provide strong evidence that SN\,2011fe marked the explosion of a white dwarf \citep{Bloom_etal12}.





Several recent papers suggest that this initial analysis may be too simplistic \citep{Piro12, Piro_Nakar12, Piro_Nakar13, Mazzali_etal13}. Since the bolometric luminosity of a SN~Ia is powered by $^{56}$Ni at early times, if the $^{56}$Ni is not mixed into the outermost regions of the ejecta, it will take some time for light to diffuse out. The diffusion time could lead to a ``dark phase" between explosion and optical rise, lasting a few hours to days depending on the radial profile of $^{56}$Ni \citep{Piro_Nakar12, Piro_Nakar13}.

Supplementing light curves with spectroscopic information about the velocity evolution of the photosphere can help account for this effect. \citet{Piro_Nakar12} use spectroscopic measurements from \citet{Parrent_etal12} to refine the explosion date of SN\,2011fe backward to UT 2011 August 23 02:30 (with a conservative uncertainly of 0.5 day). This explosion time is 14 hours earlier than that of \citet{Nugent_etal11b}. Using a different spectroscopic data set and modeling strategy, \citet{Mazzali_etal13} find an explosion time that is more than 33 hours before Nugent et al.'s estimate: UT 2011 August 22 07:00.

These results highlight the uncertainties in constraining the radius of the exploded star with shock breakout models. An earlier explosion time and longer dark phase translate into less stringent radius limits from early-time non-detections (Figure \ref{piro_nakar12fig}). If SN\,2011fe has a 24-hour long dark phase, then the photometry presented by \citet{Bloom_etal12} only limits R$_{\star}$ to $\lesssim$0.1 R$_{\odot}$ \citep{Piro_Nakar12}. As illustrated in Figure~3 of \citet{Bloom_etal12}, this less stringent limit could accommodate unusual non-degenerate stars, such as carbon or perhaps helium stars, as the exploded star in SN\,2011fe.

 \begin{figure*}[t]
\begin{center}
\includegraphics[scale=1.0, angle=0]{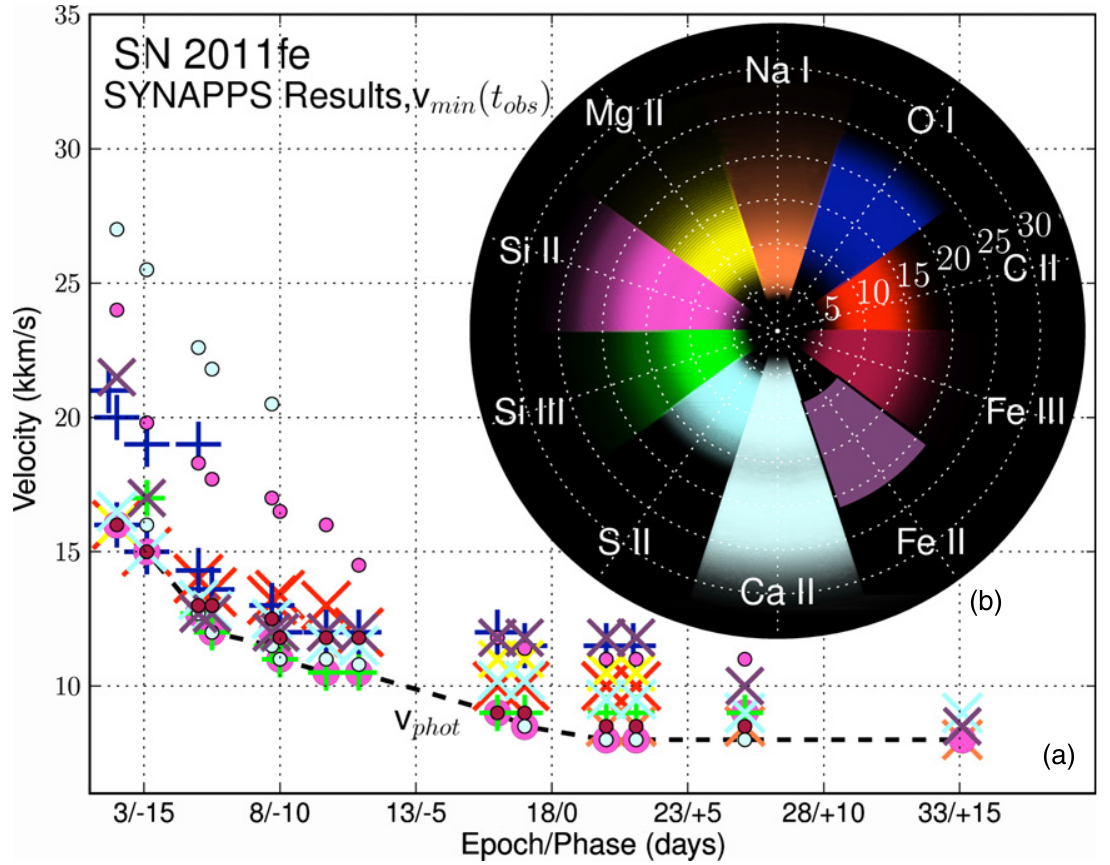}
\vspace{-0.3cm}
\caption{The distribution of ions in the ejecta of SN\,2011fe. The larger background panel (a) plots the minimum velocity measured for a given ion as a function of time. Points are color-coded to different ions as shown in the round panel (b). Panel (b) shows the velocity range observed for dominant ions. The circular white dotted lines mark radial increments of 5~000 km~s$^{-1}$, spanning ejecta velocities of 0 to 30~000 km~s$^{-1}$. Figure from \citet{Parrent_etal12}, reproduced by permission of the AAS.}\label{parrent12fig}
\end{center}
\end{figure*}

\subsection{Abundances in the exploded star}

Early-time optical spectra show significant carbon and oxygen features at a range of velocities (7~000--30~000 km s$^{-1}$) \citep{Nugent_etal11b, Parrent_etal12, Pereira_etal13, Mazzali_etal13}. Neutral carbon is also observed in IR spectra \citep{Hsiao_etal13}. SN\,2011fe is certainly not alone amongst SNe~Ia in showing carbon in its spectrum, although it is the best-studied example. In recent years, a rash of studies have found \ion{C}{ii} in many SNe~Ia spectra, provided that observations are obtained early in the explosion \citep[e.g.,][]{Parrent_etal11, Folatelli_etal12, Silverman_Filippenko12}. \citet{Nugent_etal11b} and \citet{Parrent_etal12} interpret the presence of C in early-time spectra of SN\,2011fe as evidence that carbon and oxygen are the unburnt remains of the exploded star. They conclude that the star that exploded as SN\,2011fe was likely a CO white dwarf, as commonly expected for SNe~Ia \citep[e.g.,][]{Livio01}.

 \citet{Mazzali_etal13} use observations of the fastest-moving outermost layer of ejecta ($v > $ 19~400 km s$^{-1}$) in SN\,2011fe to estimate the metallicity of the progenitor. Most of this material is carbon, but the remaining 2\% of the mass should represent the heavier elements in the progenitor white dwarf. By modeling an Fe-group absorption feature at $\sim$4800 \AA\ in conjunction with \emph{HST} UV spectroscopy, Mazzali et al.\ find a metallicity of $\sim$0.25--0.5 Z$_{\odot}$ for the outermost ejecta. \citet{Foley_Kirshner13} also use UV spectroscopy to argue that the progenitor of SN\,2011fe had sub-solar metallicity. M101's gas-phase metallicity, measured at the galactocentric radius of SN\,2011fe, is $\sim$0.5 Z$_{\odot}$ \citep{Stoll_etal11}---consistent with estimates for SN\,2011fe's progenitor system. However, it is important to keep in mind that SNe~Ia show a range of delays between the formation of the progenitor system and explosion \citep{Maoz_Mannucci12}; it would not be surprising if there was an offset between the metallicity of SN\,2011fe  and the current gas-phase metallicity in the region.
 
 The metallicity of the progenitor may be important in shaping a SN~Ia, perhaps affecting the yield of $^{56}$Ni \citep{Timmes_etal03, Jackson_etal10} and also determining the observed spectral energy distribution in the rest-frame UV, where high-redshift observations of SNe~Ia commonly take place \citep[e.g.,][]{Hoeflich_etal98, Lentz_etal00, Maguire_etal12}. The measurements in SN\,2011fe are an important data point for testing the predicted effects of metallicity on observed SNe~Ia.

\subsection{Mass of the exploded star}

SN~Ia models would be strongly constrained if we could determine whether white dwarfs must reach the Chandrasekhar mass to explode, or if sub-Chandrasekhar explosions are common. Unfortunately, estimates of the ejected mass are challenging to achieve at the necessary accuracy \citep[e.g.,][]{Mazzali_etal97, Stritzinger_etal06}. Uncertainties of $<$15\% are needed to distinguish between Chandrasekhar and sub-Chandrasekhar models, an extremely difficult task given the diversity of elements, densities, and ionic states in the ejecta of SNe~Ia.

\citet{Mazzali_etal13} estimate $\sim$1.1 M$_{\odot}$ of material is ejected at speeds $>$4~500 km s$^{-1}$ in SN\,2011fe; this determination is a model-dependent lower limit, as it does not account for the slowest moving material. Future work on nebular spectra may lead to a more complete census of the ejecta mass in SN\,2011fe \citep[e.g.,][]{Stehle_etal05}. In the meantime, the $^{56}$Ni mass places a secure lower limit on the ejecta mass in SN\,2011fe, $M_{\rm ej} > 0.5$ M$_{\odot}$.


\begin{figure*}[t]
\begin{center}
\includegraphics[scale=0.6, angle=0]{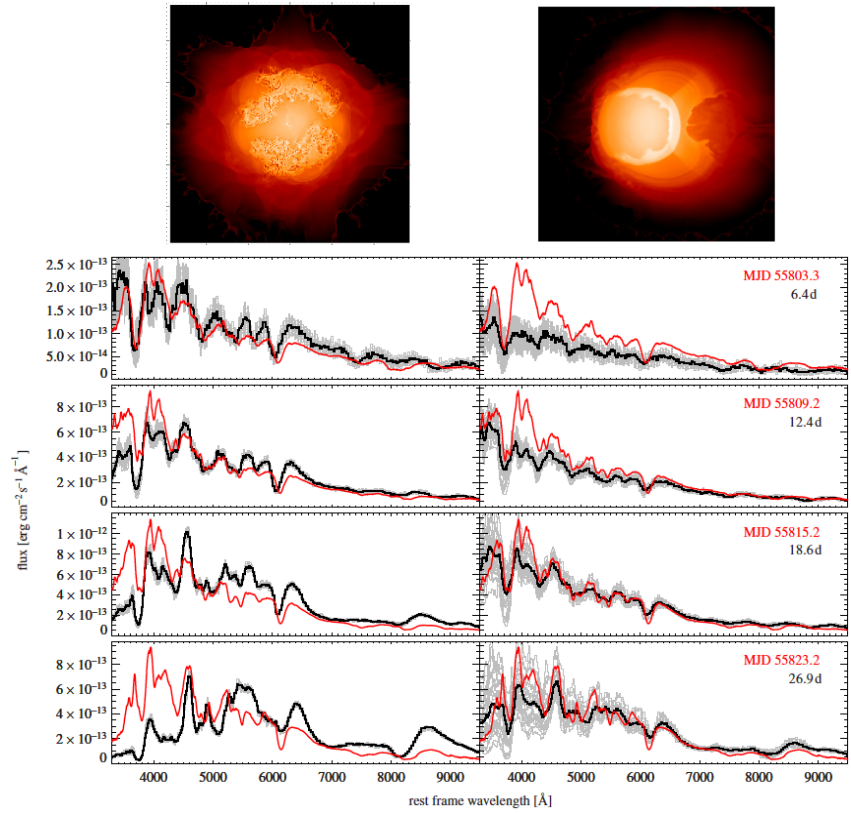}
\vspace{-1.7cm}
\caption{Top-Left Panel: Density distribution of material in a delayed detonation model 100~s after explosion. Top-right panel: Density distribution in a white dwarf merger model 100~s after explosion. Bottom Panels: Optical spectra of SN\,2011fe during four epochs spanning 6--27 days after explosion (one epoch per row). Observed spectra are shown in red, and compared with model spectra for the delayed detonation scenario (left column) and the white dwarf merger (right column). Black lines represent models averaged over all viewing angles, while grey lines show model spectra from 25 different viewing angles. Figure from \citet{Roepke_etal12}, reproduced by permission of the AAS.}\label{roepke12fig}
\end{center}
\end{figure*}

\section{How did it explode?} \label{how}

The volume and quality of data available on SN\,2011fe enable us to explore the details of the white dwarf's destruction. What is the relative significance of sub-sonic deflagration and super-sonic detonation fronts? Might the explosion have been triggered by a detonation on the white dwarf's surface? The distribution of newly-synthesized elements within the ejecta can constrain the explosion mechanism of SNe~Ia.

\citet{Parrent_etal12} obtain a time series of optical spectra and fit them using the software package \verb|SYNAPPS| in order to identify the ions contributing to each spectrum. They measure variations in velocity of each ion's features with time, and map the velocity range of each ion in Figure \ref{parrent12fig}, ranging from 5~000 to 30~000 km~s$^{-1}$. Figure \ref{parrent12fig} shows that the ejecta of SN\,2011fe are well mixed, with Si, Ca, Fe, and O present throughout much of the ejecta.  

As discussed in Section \ref{radius}, the early-time light curve contains information about the radial distribution of newly-synthesized $^{56}$Ni. By modeling the light curves and velocity evolution of SN\,2011fe, \citet{Piro12} and \citet{Piro_Nakar12} find that $^{56}$Ni must be present in the outer ejecta, constituting a mass fraction of a few percent at a mass depth of just $10^{-2}$~M$_{\odot}$ below the white dwarf's surface. Dredge-up of $^{56}$Ni to this height may present a challenge to standard delayed detonation models: these posit ignition at many points and result in relatively symmetric explosions. The observed distribution of $^{56}$Ni requires strong mixing, as might be provided by an asymmetric deflagration ignition in a delayed detonation scenario \citep{Maeda_etal10} or bubbles seen in models of gravitationally confined detonations \citep{Meakin_etal09}. However, such highly asymmetric models generally conflict with observations of SNe~Ia \citep{Blondin_etal11} and with spectropolarimetric observations of SN\,2011fe \citep[][see below for more discussion]{Smith_etal11}. Alternatively, a double detonation scenario (where a He-rich shell detonates on the surface of the white dwarf and drives a shock inward, inducing nuclear burning of the entire white dwarf) might also explain the presence of Fe-group elements at the outer edges of the ejecta \citep{Piro_Nakar12}. However, double detonation models also struggle to match the observed spectra of SNe Ia \citep{Kromer_etal10, Woosley_Kasen11}. Recent modeling of the standard delayed-detonation scenario can yield $^{56}$Ni at large radii along some lines of sight \citep{Seitenzahl_etal13}; more work is required to determine if such models can explain the observations of SN\,2011fe.

\citet{Mazzali_etal13} compare a time series of UV+optical spectra with SN~Ia explosion models. They consider a pure deflagration model in the form of the famous benchmark W7 \citep{Nomoto_etal84}, which has a steep density profile at the outermost radii and very little mass expanding at the highest velocities. This model produces good fits to optical spectra, but it overpredicts the flux in the UV---more material is required at large velocity, above the photosphere, in order to absorb this light. The W7 model is contrasted with a delayed detonation model \citep{Iwamoto_etal99}, which has significantly more material expanding at high velocities ($>$16~000 km~s$^{-1}$). However, this model predicts larger blueshifts to the UV Fe-group features than observed. Therefore, \citet{Mazzali_etal13} compose a hybrid model with an outer density profile of intermediate steepness between the pure-deflagration and delayed detonation models. This hybrid essentially corresponds to a weak delayed detonation and provides a better fit to the optical+UV spectra.

Observations of SN\,2011fe are also compared with two different three-dimensional explosion models by \citet{Roepke_etal12}. One model is for a delayed detonation of a Chandrasekhar-mass white dwarf, while the other represents a violent merger of two WDs (1.1 M$_{\odot}$ + 0.9 M$_{\odot}$). Both models produce the right amount of $^{56}$Ni to match the light curves of SN\,2011fe, and both models can fit the spectra of SN\,2011fe reasonably well (Figure \ref{roepke12fig}). The delayed detonation model matches the early time spectra better, while the merger provides a significantly better fit at later times. In both models, the predicted spectra are blue-shifted relative to the data on SN\,2011fe; this discrepancy might be resolved by increasing the oxygen-to-carbon ratio in the progenitor white dwarf (assumed in the R{\"o}pke et al. study to be 1:1).

Each model of \citet{Roepke_etal12} provides a range of spectra, varying with viewing angle because the model explosions are not spherically symmetric (grey lines in Figure \ref{roepke12fig}). The merger is significantly more asymmetric than the delayed detonation (top row of Figure \ref{roepke12fig}), so the spectra predicted from the merger model cover a wider swath of possible observations. Future work is needed to test if the relatively asymmetric explosions predicted by white dwarf mergers are inconsistent with spectropolarimetric observations of SNe~Ia, which constrain the geometry of the ejecta \citep{Wang_Wheeler08}.

The spectropolarimetric observations of \citet{Smith_etal11} find that SN\,2011fe is polarized at a level of only $\sim$0.2--0.4\%. However, compared with the continuum and other spectral lines, the strong \ion{Si}{ii}$\lambda$6755 feature has different time-dependent polarization properties. Smith et al.~propose a geometric model wherein the continuum photosphere is an ellipse elongated in the polar direction, with a Si-rich ``belt" stretching along the equator. While a unique interpretation of the spectropolarimetry of SN\,2011fe is difficult, the observations hint at some small departures from spherical symmetry.

Most constraints therefore imply that SN\,2011fe is consistent with a mildly asymmetric delayed detonation model.
In the future, the late-time light curve of SN\,2011fe ($\gtrsim$4 years after explosion) may distinguish between explosion models \citep{Roepke_etal12}. The amount of radioactive $^{55}$Fe in the ejecta (half life: 2.75 yr) scales with the central density of the exploded white dwarf. Therefore, the delayed detonation of a Chandrasekhar-mass white dwarf should be brighter at late times than the merger or two sub-Chandrasekhar white dwarfs. However, it will be challenging to infer the bolometric luminosity solely from optical photometry \citep{McClelland_etal13}, and the effect of $^{55}$Fe will need to be carefully disentangled from the possible late-time contribution of the puffed-up companion star (see Section \ref{progen}; \citealt{Shappee_etal13a}).

\section{What is the progenitor of SN\,2011fe?} \label{progen}

It is generally thought that, in order to explode as a SN~Ia, a CO white dwarf must be destabilized by mass transfer from a binary companion. However, the nature of the companion---whether a main sequence, subgiant, or giant star in a ``single-degenerate" binary, or another white dwarf in a ``double-degenerate" binary---remains unknown. Similar uncertainties exist for the nature of the mass transfer, whether gradual accretion or a sudden merger.



\begin{figure}[h]
\begin{center}
\includegraphics[scale=0.67, angle=0]{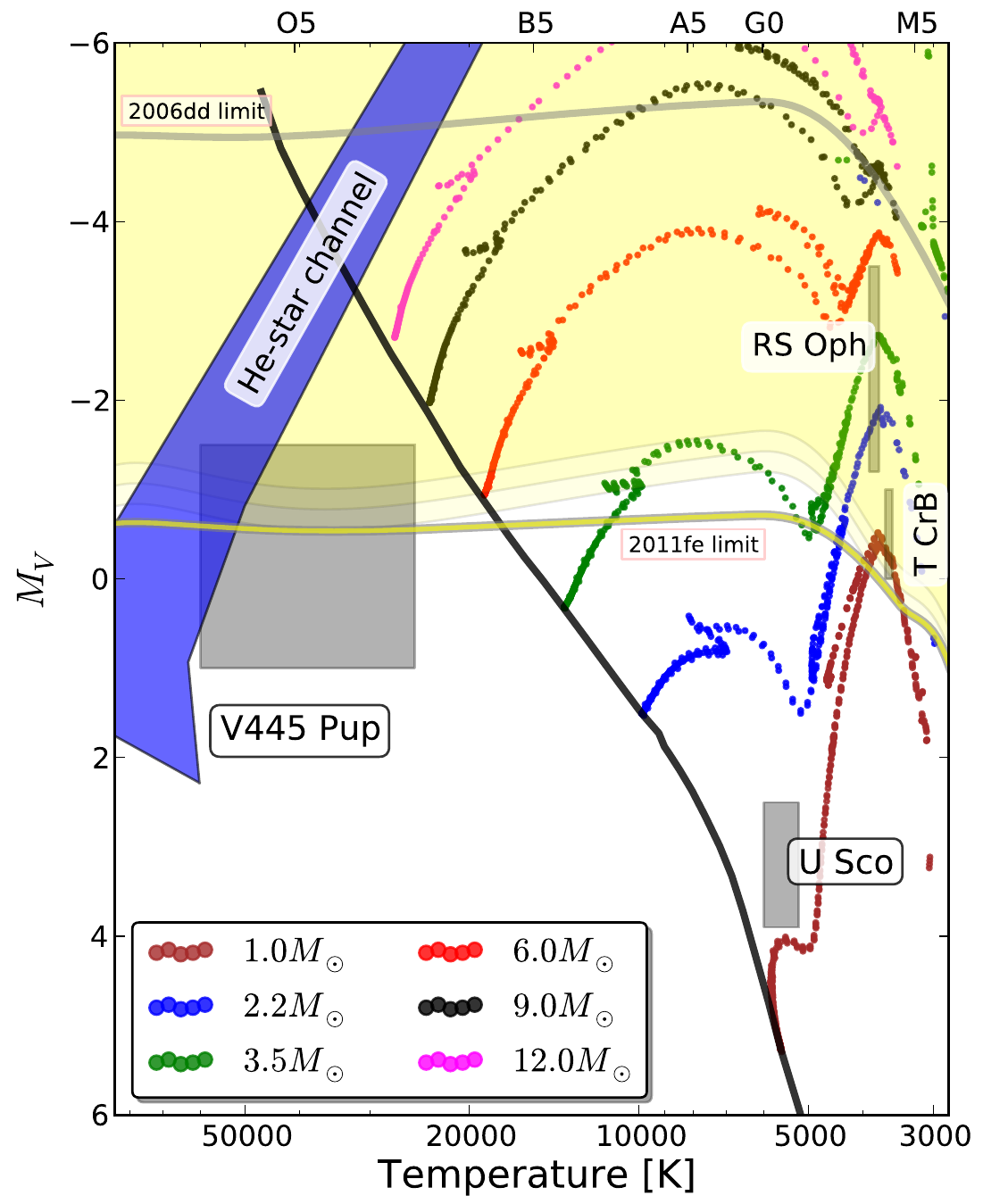}
\vspace{-0.4cm}
\caption{A Hertzprung-Russell diagram showing the limits on a companion star to SN\,2011fe, derived from pre-explosion \emph{HST} imaging. The parameter space above the yellow line is ruled out, excluding most red giants as the companions to SN\,2011fe. The stellar main sequence is plotted as a black line and giant branches for stars of various masses are plotted as colored dots (key in bottom left). Several famous candidates for single-degenerate SN~Ia progenitors are also plotted as shaded grey regions: recurrent novae with red giant companions (RS~Oph and T~CrB), a recurrent nova with a main-sequence companion (U~Sco), and a He nova (V445~Pup). Figure from \citet{Li_etal11b}. Reprinted by permission from Macmillan Publishers Ltd: Nature, copyright 2011.}\label{li12fig}
\end{center}
\end{figure}

\subsection{Constraints on the companion star}
\citet{Li_etal11b} study deep pre-explosion \emph{HST} imaging of the site of SN\,2011fe in M101. No source is present at the location of SN\,2011fe, ruling out bright binary companions like most red giants (Figure \ref{li12fig}). These are by far the deepest pre-explosion limits placed on the progenitor of a SN~Ia; others have been factors of $>$60 less sensitive and also yielded non-detections \citep{Maoz_Mannucci08}. Still, Li et al.~cannot exclude main sequence or subgiant companions of $\lesssim$4 M$_{\odot}$.

If the SN shock plows over a non-degenerate companion star, this interaction is expected to produce an early-time blue ``bump" in the UV/optical light curve \citep{Kasen10}. The amplitude of this bump depends on the binary separation and viewing angle. \citet{Brown_etal12} find no such feature in \emph{Swift}/UVOT photometry of SN\,2011fe (see also \citealt{Bloom_etal12}; Figure \ref{bloom12fig}), and constrain the binary separation to $\sim$few $\times 10^{11}$ cm ($\sim$0.01 AU). This constraint also rules out red giant companions, and, assuming mass transfer by Roche Lobe overflow, it implies a mass of  $\lesssim$1 M$_{\odot}$ for a potential main sequence companion. However, the dark phase predicted by \citet{Piro_Nakar12} and discussed in Section \ref{how} may soften these constraints by a factor of several; future work is needed to self-consistently model the dark phase and the ejecta's interaction with a companion.

\citet{Shappee_etal13b} also place constraints on the companion star to SN\,2011fe by searching for H$\alpha$ emission in a nebular spectrum nine months after explosion. If the companion to a SN~Ia is non-degenerate, $\sim$0.1--0.2 M$_{\odot}$ of hydrogen is predicted to be swept from the companion and entrained in the low-velocity ejecta \citep{Marietta_etal00, Pan_etal12, Liu_etal12}. Once the ejecta become optically thin, the hydrogen-rich material should be observable as H$\alpha$ emission \citep{Mattila_etal05}. Studies of previous SNe~Ia constrained the entrained H to $\lesssim$0.01 M$_{\odot}$ \citep{Leonard07}, but \citet{Shappee_etal13b} place an order-of-magnitude stronger limit in SN\,2011fe, $\lesssim$0.001 M$_{\odot}$. If this result holds, it would essentially exclude all non-degenerate secondaries and require a double-degenerate model for SN\,2011fe. However, more theoretical work is needed to thoroughly explore gamma-ray trapping in the ejecta, which is responsible for powering the H$\alpha$ emission. Considerable uncertainties remain in predicting the H$\alpha$ luminosity associated with a given mass of entrained hydrogen, but late-time H$\alpha$ observations hold promise for constraining the companions of SNe~Ia.

A final test of the companion to SN\,2011fe is possible from late-time observations of the light curve. A non-degenerate companion should expand and grow in luminosity after being shocked by the supernova blast wave; it is expected to remain a factor of $10-10^{3}$ over-luminous for $\sim$10$^{3}-10^{4}$ yr \citep{Shappee_etal13a}. The signature of such a puffed-up companion should be visible in SN\,2011fe $\gtrsim$3.5 years after explosion, but could at first be confused with variations in radioactive yields from the SN itself \citep{Roepke_etal12}. Very late time measurements will effectively search for such an altered companion at the site of SN\,2011fe.

\subsection{Constraints on the circumbinary medium}
A red giant progenitor is also ruled out by searches for circumbinary material using radio and X-ray observations \citep{Horesh_etal11, Chomiuk_etal12, Margutti_etal12}. The interaction between a supernova blastwave and the circumstellar medium accelerates particles to relativistic speeds and amplifies the magnetic field in the shock front, producing radio synchrotron emission \citep{Chevalier82b, Chevalier98}. These same relativistic electrons also up-scatter photons radiated by the supernova itself, producing inverse Compton radiation at X-ray wavelengths \citep{cf06}. While these signals are often observed in nearby core-collapse SNe \citep{Weiler_etal02, Soderberg_etal06b}, no SN~Ia has ever been detected at radio or X-ray wavelengths, implying that SNe~Ia do not explode in dense environments \citep{Panagia_etal06, Immler_etal06, Hancock_etal11, Russell_Immler12}. SN\,2011fe is no exception, with multiple epochs of non-detections in deep radio and X-ray data.

With SN\,2011fe, we can place the most stringent limits to date on the circumbinary environment around a SN~Ia. Assuming the circumbinary material is distributed in a wind profile ($\rho = {{\dot{M}}\over{4 \pi v_w}}  r^{-2}$, where $\dot{M}$ and $v_w$ are the mass-loss rate and velocity of the wind), \citet{Chomiuk_etal12} use deep radio limits from the Karl G.~Jansky Very Large Array to find that $\dot{M} \lesssim 6 \times 10^{-10}$~M$_{\odot}$~yr$^{-1}$ in the surroundings of SN\,2011fe, for $v_w = 100$ km s$^{-1}$. Assuming a uniform density medium, they find its density must be $n_{\rm CSM} \lesssim 6$~cm$^{-3}$. These limits on the circumbinary medium not only rule out a red giant companion for SN\,2011fe, but also exclude optically-thick accretion winds and non-conservative mass transfer during Roche Lobe overflow \citep{Chomiuk_etal12}. The environment around SN\,2011fe is extremely low-density.
 
 \citet{Horesh_etal11} caution that radio limits on the circumbinary density depend on poorly-understood microphysical parameters governing the efficiency of particle acceleration and magnetic field amplification. \citet{Margutti_etal12} use X-ray observations from \emph{Chandra} and \emph{Swift} to place constraints on the circumbinary medium which are less model-dependent than the radio limits, because they do not depend on an assumed magnetic field strength. While the X-ray limits on circumbinary density are somewhat less stringent than the radio constraints, it is worth noting that the sensitivity of X-ray observations to circumstellar material scales with the bolometric luminosity of the supernova. The deep \emph{Chandra} observation on SN\,2011fe was obtained just three days after discovery, significantly before light curve peak. If instead \emph{Chandra} had observed at optical maximum, the X-ray constraints on circumbinary material around SN\,2011fe would be more stringent than the radio limits. 
 
A search for circumstellar dust carried out by \citet{Johansson_etal13} uses imaging from \emph{Herschel} at 70~$\mu$m and 160~$\mu$m. Both pre- and post-explosion imaging yield non-detections, constraining the dust mass in the vicinity of SN\,2011fe to $\lesssim$ 7$\times 10^{-3}$~M$_{\odot}$ (assuming a dust temperature of 500~K).
 
A  clean circumbinary environment is also found by \citet{Patat_etal13}, who study optical absorption lines along the line of sight to SN\,2011fe.  In a few SNe~Ia, time-variable \ion{Na}{i}~D absorption has been observed and attributed to the presence of circumbinary material that is ionized by the SN and then recombines \citep{Patat_etal07, Simon_etal09}. \citet{Patat_etal13} obtain multi-epoch high-resolution spectroscopy of SN\,2011fe and find no evidence for time variability in the \ion{Na}{i}~D profile, again implying a lack of significant circumbinary material.

\subsection{Constraints on the accretion history}
The above-described constraints rule out many single-degenerate progenitors, and are often cited as evidence that SN\,2011fe was the product of a white dwarf merger. However, the constraints can not conclusively exclude a main-sequence or sub-giant donor of reasonably low mass, $\lesssim$1--2~M$_{\odot}$, transferring material via Roche lobe overflow. 

At low mass transfer rates, such a system might look like the recurrent nova U~Sco (Figure \ref{li12fig}), which ejects $10^{-6}$~M$_{\odot}$ every $\sim$10 years in a nova explosion and harbors a white dwarf near the Chandrasekhar mass (\citealt{Thoroughgood_etal01, Diaz_etal10, Schaefer10}; although the white dwarf in U~Sco is likely composed of ONe, rather than CO; \citealt{Mason_etal11}). \citet{Li_etal11b} collect a time-series of pre-explosion photometry at the site of SN\,2011fe to search for novae preceding the supernova. All epochs yield non-detections, and they estimate a $\sim$60\% chance that a nova would have been detected if it had erupted in the five years prior to SN\,2011fe. There have also been suggestions in the literature that nova shells around SNe~Ia should produce time-variable NaD absorption features \citep{Patat_etal11}; \citet{Patat_etal13} find no evidence of a nova-like shell surrounding SN\,2011fe.

In a progenitor system with a main sequence companion transferring mass at higher rates, steady burning of hydrogen is expected on the white dwarf surface, instead of unstable burning in the form of novae. The white dwarf will radiate thermal emission of $\sim$few~$\times 10^{5}$ K, with a spectral energy distribution peaking in the far-UV to soft X-ray. \citet{Liu_etal11} and \citet{Nielsen_etal12} search deep pre-explosion X-ray imaging of the site of SN\,2011fe, looking for evidence of such a super-soft X-ray source, and emerge with non-detections. While this constraint rules out many known super-soft sources, it is not stringent enough to exclude those with lower luminosities or cooler temperatures; for example, a source like the persistent super-soft source Cal~83 remains viable \citep{Liu_etal11}.

One piece of evidence suggesting a single-degenerate system is that the fastest-moving ejecta ($>$19~400~km~s$^{-1}$) in SN\,2011fe are almost exclusively composed of carbon (98\% by mass; \citealt{Mazzali_etal13}). Mazzali et al.\ interpret these outermost ejecta as the ashes of the material accreted onto the white dwarf before the SN. If SN\,2011fe marked the merger of two CO white dwarfs, a significant fraction of this material should be oxygen---but the O fraction is small and strongly constrained by the observed \ion{O}{i} feature at 7774 \AA. They conclude that the dominance of C in highest-velocity ejecta is support for H-rich accretion onto the white dwarf, because H will fuse to C on the outskirts of a SN~Ia, but the ejecta will expand before significant amounts of C can subsequently fuse to O. The outer ejecta might also be consistent with the accretion of helium under special circumstances, but in most conditions He should burn explosively up to the Fe-peak.

A relatively exotic strategy for ``hiding" the non-degenerate companion of a SN~Ia was proposed by \citet{Justham11} and \citet{diStefano_etal11} and dubbed the ``spin-up/spin-down" model. A white dwarf accreting from a non-degenerate companion may reach significant rotational speeds by conservation of angular momentum, and centripetal force will help support the white dwarf and prevent it from exploding as a SN~Ia. Upon the cessation of mass transfer (presumably due to the evolution of the companion), the white dwarf will begin to spin down, and after a delay, it will finally explode as a SN~Ia. The spin-down time is uncertain and potentially highly variable, $\sim$10$^{3}-10^{10}$~yr. This delay may provide sufficient time for the evolved companion to lose any remaining H-rich envelope and contract to a small and unobtrusive radius. While this model can reconcile SN\,2011fe to a range of single-degenerate progenitor systems \citep{Hachisu_etal12}, it is highly speculative. Accreting white dwarfs are observed to spin at significantly lower rates than predicted by simple conservation of momentum \citep{Sion99}, and the models of spinning white dwarfs remain preliminary \citep[e.g.,][]{Yoon_Langer05}.

\begin{figure*}[t]
\begin{center}
\includegraphics[scale=0.7, angle=0]{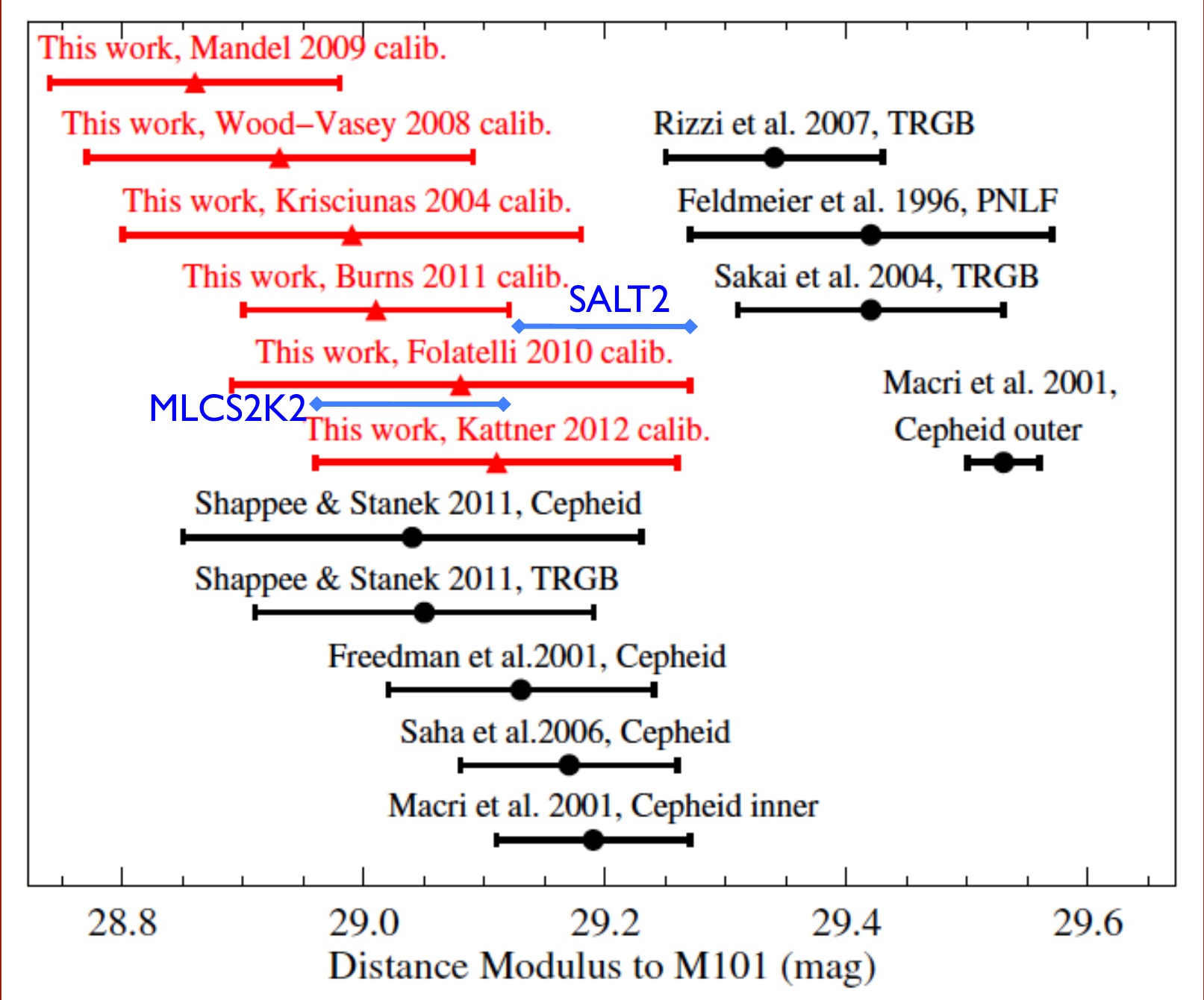}
\vspace{-0.1cm}
\caption{Distance moduli measured to M101 with a range of techniques and 1$\sigma$ error bars. Black lines are pre-SN\,2011fe estimates which use Cepheids, TRGB, or PNLF methods. Red lines use near-IR light curves of SN\,2011fe to estimate the distance, and represent fits to a variety of calibrations \citep{Matheson_etal12}. Blue lines use $BVRI$ light curves of SN\,2011fe with the fitters SALT2 and MLCS2k2 \citep{Vinko_etal12}. Figure modified from \citet{Matheson_etal12} to include data from \citet{Vinko_etal12}; reproduced by permission of the AAS.}\label{matheson12fig}
\end{center}
\end{figure*}
 
\section{How accurate are SNe~Ia as standardizeable candles?}

Because the distance to M101 is relatively well known and because SN\,2011fe is so well-studied, it constitutes a test of SNe~Ia as standardizeable candles. \citet{Matheson_etal12} collect distance measurements to M101 obtained independent of SN\,2011fe, using Cepheid variable stars \citep{Freedman_etal01, Macri_etal01, Saha_etal06, Shappee_Stanek11}, the tip of the red giant branch (TRGB; \citealt{Sakai_etal04, Rizzi_etal07, Shappee_Stanek11}; also \citealt{Lee_Jang12}), and the planetary nebula luminosity function (PNLF; \citealt{Feldmeier_etal96}). These estimates span distance moduli of 29.04--29.53 (Figure \ref{matheson12fig}) with a standard deviation of 0.18 mag.

The light curves of SN\,2011fe can be used to independently estimate the distance to its host galaxy. \citet{Vinko_etal12} observe optical light curves in $BVRI$ and apply two often-used light curve fitters to estimate the distance to SN\,2011fe: MLCS2k2 \citep{Jha_etal07} and SALT2 (Figure \ref{vinko12fig}; \citealt{Guy_etal07, Guy_etal10}). The best-fit MLSCS2k2 template returns a distance modulus of $29.21 \pm 0.07$ mag, while SALT2 yields $29.05\pm0.08$ mag (Figure \ref{matheson12fig}); both assume $H_{0} = 73$ km s$^{-1}$ Mpc$^{-1}$. The errors in distance moduli are dominated by degeneracies in the template fits, not by noise in the data. The difference in distance modulus measured from these two calibrations is consistent with the scatter in each calibration measured from a larger sample of SNe~Ia ($\sim$0.15 mag; \citealt{Kessler_etal09}).

The scatter observed between SN~Ia light curves is smaller in the near-IR than in the optical \citep{Phillips11}. Therefore, \citet{Matheson_etal12} carry out a similar procedure as \citet{Vinko_etal12}, but use $JHK_{s}$ light curves of SN\,2011fe to compare various SN~Ia calibrations. Plotted in red in Figure \ref{matheson12fig} are distance moduli calculated using the $H$-band peak apparent brightness of SN\,2011fe and six different calibrations of near-IR light curves (\citealt{Krisciunas_etal04, Wood-Vasey_etal08, Mandel_etal09, Folatelli_etal10, Burns_etal11, Kattner_etal12}; assuming $H_{0} = 72$ km s$^{-1}$ Mpc$^{-1}$). The calibrations yield distance moduli ranging from 28.93 mag to 29.17 mag. Each measurement has a significant uncertainty associated with it, $\sim$0.16 mag, due to the scatter in the data used to develop the calibration. The distribution of distance moduli calculated from these calibrations has a standard deviation of 0.12 mag, consistent with the error in a single calibration.

\citet{Vinko_etal12} estimate that the error on the distance to M101, using estimates from both Cepheids and SN\,2011fe, remains at $\sim$8\%----a rather large uncertainty, given that M101 is one of the best-studied nearby galaxies. Still, distance estimates to SN\,2011fe agree with recent Cepheid and TRGB distance determinations to M101 within 1$\sigma$ of quoted systematic errors on these calibrations. These results imply that the systematic errors in standard candle calibrations are well-estimated and large offsets do not exist in zero points.

\section{Conclusions}

\noindent $\bullet$ \textbf{What exploded in SN\,2011fe?} A carbon/oxygen white dwarf of sub-solar metallicity.

\noindent $\bullet$ \textbf{How did it explode?} Most data are consistent with a slightly asymmetric delayed detonation, but other scenarios might also fit, if studied in more detail.

\noindent $\bullet$ \textbf{What is the progenitor of SN\,2011fe?} Despite much deeper searches than in any preceding SN~Ia, very little evidence for a non-degenerate companion is found in SN\,2011fe. Small corners of single-degenerate parameter space remain viable, but the data imply that SN\,2011fe may have been the merger of two white dwarfs. It is important to remember that SN\,2011fe is just one supernova, and the class of SNe~Ia may be diverse. 

\noindent $\bullet$ \textbf{How accurate are SNe~Ia as standardizeable candles?} Different calibrations of SN~Ia light curves, when applied to SN\,2011fe, yield a range of distances to M101 with a standard deviation of 11\%. These agree with Cepheid and TRGB distances to M101 at the 1$\sigma$ level.

Because of its early discovery, proximity, and normalcy, SN\,2011fe constitutes a unique opportunity for detailed study of a SN~Ia. It is likely to be a decade or more before the next similarly bright and nearby SN~Ia explodes, and in the meantime theorists should continue to develop models and revisit the exquisite data collected for SN\,2011fe, with the goal of further constraining its progenitor system and explosion mechanism. For example, additional work is needed to accurately predict the H$\alpha$ luminosity expected from a single-degenerate SN Ia in the nebular phase. Spectra of SN\,2011fe, spanning just one day after explosion to the late nebular phase and the UV to the IR, are a rich observational resource which have just begun to be tapped. Papers 
modeling the spectra have, to date, considered only a small handful of specific explosion models; future work should more thoroughly explore the parameter space of plausible explosion mechanisms, analyze the uniqueness of predicted observables from different models, and consider all observables when comparing with models. 

Late-time photometry on SN\,2011fe should be pursued for years to come, with goals of constraining yields of radioactive isotopes and searching for a puffed-up companion star.

Continued efforts to compare observations of SN\,2011fe with theory will ensure a solid groundwork for interpreting the large samples of SNe~Ia to be obtained with LSST. When the next nearby bright SN~Ia explodes, we will be in an even better position to test models of SNe~Ia, armed with the next generation of time-domain telescopes like LSST, ASKAP, and LOFT and a polished theoretical framework. 

\section*{Acknowledgments} 
L.~C. is a Jansky Fellow of the National Radio Astronomy Observatory. She thanks the organizers of the ``Supernovae Illuminating the Universe: From Individuals to Populations" Conference, which took place September 2012 in Garching, Germany. Their kind invitation for a review talk on SN\,2011fe was the inspiration for this article. She is also grateful to Ken Shen, Fernando Patat, Alicia Soderberg, Max Moe, and Dean Townsley for helpful discussions, Stuart Ryder and Bryan Gaensler for their patience and work, and an anonymous referee for useful comments.

\bibliographystyle{apj}


\bibliography{refs}












\end{document}